\documentclass[onecolumn,showpacs,showkeys,floatfix,aps,nofootinbib,11pt]{revtex4}
\usepackage{amssymb,amsmath}
\usepackage[dvips]{graphics,color}
\usepackage{epsfig}\usepackage{float}
\newcommand{\beq}{\begin{eqnarray}}
\newcommand{\eeq}{\end{eqnarray}}

\newcommand{\ba}{\begin{eqnarray}}
\newcommand{\ea}{\end{eqnarray}}
\newcommand{\bege}{\begin{equation}}
\newcommand{\enge}{\end{equation}}
\newcommand{\benu}{\begin{enumerate}}
\newcommand{\enu}{\end{enumerate}}

\newcommand{\noi}{\noindent}
\newcommand{\bbbbox}{\mathop{\Box\kern -5pt\raisebox{.8pt}{$|$}}}

\newcommand{\RR}{\mathbb{R}}
\newcommand{\mt}{\mathcal}
\newcommand{\cl}{\mt{C}\ell}

\newcommand{\la}{\Lambda}

\newcommand{\w}{\wedge}
\newcommand{\vv}{{\bf v}}

\newcommand{\OO}{\mathbb{O}}

\def\beq{\begin{eqnarray}}
\def\eeq{\end{eqnarray}}

\def\0{\mbox{\boldmath$\displaystyle\mathbf{0}$}}

\def\h00h{\mbox{\boldmath$\displaystyle\mathbf{(1/2,0)\oplus(0,1/2)}$}}

\def\ba{p \,e^{i\phi} \sin(\theta)}

\newfont{\gotico}{eufm10 scaled\magstephalf}
\newfont{\qvd}{msam10 scaled\magstephalf}

\def\de#1/de#2{\frac{\partial {#1}}{\partial {#2}}}
\def\De#1/de#2{\dfrac{\partial {#1}}{\partial {#2}}}

\begin{document}
\vskip-2cm
\title{Flag-Dipole Spinor Fields in ESK Gravities}
\author{Rold\~ao da Rocha} \email{roldao.rocha@ufabc.edu.br} \affiliation{Centro de Matem\'atica, Computa\c c\~ao e Cogni\c c\~ao, Universidade Federal do ABC, 09210-170, Santo Andr\'e, SP, Brazil}
\author{Luca Fabbri} \email{fabbri@diptem.unige.it} \affiliation{DIPTEM Sez. Metodi e Modelli Matematici, Universit\`{a} di Genova, Piazzale Kennedy, Pad. D - 16129 Genova and INFN \& Dipartimento di Fisica, Universit\`{a} di Bologna, Via Irnerio 46, - 40126 Bologna, Italia}\author{J. M. Hoff da Silva} \email{hoff@feg.unesp.br; hoff@ift.unesp.br} \affiliation{UNESP - Campus de Guaratinguet\'a - DFQ Av. Dr. Ariberto Pereira da Cunha, 333, 12516-410, Guaratinguet\'a - SP, Brazil}
\author{R. T. Cavalcanti} \email{rogerio.cavalcanti@ufabc.edu.br} \affiliation{CCCCNH, Universidade Federal do ABC, 09210-170, Santo Andr\'e, SP, Brazil}
\author{J. A. Silva-Neto} \email{jose.antonio@ufabc.edu.br} \affiliation{CCCCNH, Universidade Federal do ABC, 09210-170, Santo Andr\'e, SP, Brazil}
\begin{abstract}
\noindent We consider the Riemann-Cartan geometry as a basis for the Einstein-Sciama-Kibble theory coupled to  spinor fields: we focus on $f(R)$ and conformal gravities, regarding  the flag-dipole spinor fields, type-(4) spinor fields under  the Lounesto classification. We study such theories in specific cases given for instance by cosmological scenarios: we find that in such background the Dirac equation admits solutions that are not Dirac spinor fields, but in fact the aforementioned flag-dipoles ones. These solutions are important from a theoretical perspective, as they evince  that spinor fields are not necessarily determined by their dynamics, but also a discussion on their structural (algebraic) properties must be carried off. Furthermore, the phenomenological point of view is shown to be also relevant, since for  isotropic Universes they circumvent the question whether spinor fields do undergo the Cosmological Principle.
\end{abstract} 
\maketitle
\section{Introduction}
In a geometry that incorporates a differential structure, the introduction of covariant derivatives is as inevitable as the definition of the metric. Moreover, the connection in the most general case is not symmetric as well as the metric is not a constant, giving rise respectively to torsion and curvature.

On the other hand, torsion might play an important role from a genuine physical point of view, as the spacetime curvature already does and is undeniably measured in numerous experiments. In fact, according to the Wigner classification of particles in terms of their masses and spin, physical fields are known to be characterized by both the energy and the spin density. In the most general case, all geometric quantities can be coupled to corresponding physical fields, through specific field equations. Therefore, in the same spirit in which Einstein gravity couples curvature to energy, in the most general case this coupling is still valid. Besides, it is also accompanied by a correspondent coupling between torsion and spin. Then Einstein gravity is not the most general case but the most general spinless situation, in the sense that it is only the most general dynamical solution in absence of any spinning matter of Einstein-Sciama-Kibble theory \cite{h-h-k-n}. Here, by Einstein-Sciama-Kibble (ESK) theory we assume it in a broad sense as any torsional completion of gravity, no matter the order-derivatives of the field equations that define it.

The structure of ESK gravity is then constructed on the scheme for which we have curvature-energy as well as torsion-spin field equations. What is known to be the ESK theory in the strict sense is realized by insisting that those field equations have the least-order derivative possible, but more general ESK-like theories are possible by relaxing this condition. Of course all possible ESK-like theories are infinite, not all of them are physical, and even among the physical ones, not all of them can be sensibly considered. So a choice is to be made, and ours shall be on those that at the moment are the most in fashion: $f(R)$-gravity and the conformally gravity.

Motivated by those considerations, the ESK theory may be considered, both in its $f(R)$ and in its conformal realization, with general spinning matter fields. For the case of spinorial matter, throughout the paper we shall employ spin-$\frac{1}{2}$ spinor fields, solely. As for the case of gravity we allowed ourselves to consider higher-order derivative extensions, for the case of matter fields we shall do the same by allowing ourselves to take into account higher-order derivative spinorial matter field equations, but we shall not take into account higher-spin fields. This restriction can be dictated by the fact that such higher-spin fields may be unphysical, displaying inconsistency, non-causality, and other problems \cite{v-z}, or simply because it is not possible to consider all possibilities and a choice must be made. 
Nevertheless, as just noticed, we shall allow ourselves to go beyond the first-order derivative field equations.

In this paper, the  spinor fields shall not be called Dirac spinors, as many more possibilities can be met \cite{lou2}. In \cite{lou2}, Lounesto proceeds with the classification of the possible spin-$\frac{1}{2}$ spinors, categorizing them within six classes: Dirac fields, in various forms, belong to the first three of them; flag-dipoles and flagpoles are the fourth and fifth type of spinor fields, disseminated in the literature as a mathematical apparatus to support Penrose flags \cite{flags}, among other interesting applications; Weyl spinors is within the sixth class. As the first three classes, as well as the fifth and sixth ones, are prominently relevant in quantum field theory and its phenomenology, the fourth class should be better understood, and hereon we shall therefore focus mainly on the flag-dipoles.

No matter what spinor field we consider, in ESK theories torsion shall always be coupled to the spin density of the matter field. Therefore, after that all terms involving the covariant derivatives and the curvature are split in their torsionless counterparts plus torsional contributions, the latter can be substituted through the torsion-spin field equations in terms of the spin density of the spinorial matter field. All field equations of the ESK theory thus reduce to the same equations of the torsionless theory complemented by spin-spin self-interacting potentials, and thus non-linearities appear in the matter field equations. This is general, and the specific gravitational background ($f(R)$ of conformal) and type of spinor (Dirac or flag-dipole, or other still) shall determine the exact structure of these non-linearities in the matter field equations. For example, in the least-order derivative ESK gravity with Dirac fields, the non-linearities are given in terms of axial current squared contact interactions, that is with the structure of the Nambu-Jona-Lasinio (NJL) potential \cite{n-j--l/1}. As we shall see, in $f(R)$ gravity they shall turn out to be structurally similar apart from a scaling function as a running coupling, while in conformal gravity they might be entirely different. In all these cases however, when the spinor field is a flag-dipole, the interaction is shown to change.

In what follows, we aim to study the flag-dipole type-(4) spinor fields dynamics in the case of an ESK theory, whether $f(R)$ or conformal gravity: we plan to show that in such a context, matter fields which are solutions of the Dirac equation are not necessarily Dirac spinor fields by exhibiting a physical solution of the Dirac equation that is instead a flag-dipole spinor field; in this case then, we shall be able to show, through a specific example, that a spinor field is not fully determined by its dynamics since spinor fields obeying the Dirac equation are not necessarily Dirac spinor fields.

This paper is organized as follows: in the next Section we introduce the Lounesto classification program according to the bilinear covariants and provide some necessary concepts concerning type-(4) spinor fields. In Section III we study the spinor fields solutions in the context of torsional $f(R)$ gravity and conformal gravity cases, showing that they are non-standard singular classes under Lounesto spinor field classification. In Section IV we conclude. In the Appendix we show how to construct the most general type-(4) flag-dipole spinor field.
\section{Non-Standard (Flag-dipole) Spinor Fields}
This Section is devoted to briefly provide some properties on the flag-dipole spinor fields, where the most relevant general properties regarding such spinor fields, and the notation fixed throughout the text as well, are introduced.

Classical spinor fields carry the $(1/2,0)\oplus {(0,1/2)}$ representation of the Lorentz group SL$(2,\mathbb{C)\simeq }\;\,\mathrm{Spin}_{1,3}^{e}$. They are sections of the vector bundle $\mathbf{P}_{\mathrm{Spin}_{1,3}^{e}}(M)\times _{\rho }\mathbb{C}^{4}, $ where $\rho $ denotes the ${(1/2,0)}\oplus {(0,1/2)}$
representation of SL$(2,\mathbb{C})$ in $\mathbb{C}^{4}$. Furthermore, classical spinor fields can be sections of the vector bundle $\mathbf{P}_{\mathrm{Spin}_{1,3}^{e}}(M)\times _{\rho ^{\prime }}\mathbb{C}^{2},$ where $\rho ^{\prime }$ is the ${(1/2,0)}$ or the ${(0,1/2)}$ representation of SL$(2,\mathbb{C})$ in $\mathbb{C}^{2}$. Given a spinor field $\psi$, the bilinear covariants are defined as:\begin{align}
\sigma & =\psi ^{\dagger }\gamma _{0}\psi ,\quad \mathbf{J}=J_{\mu }\theta
^{\mu }=\psi ^{\dagger }\gamma _{0}\gamma _{\mu }\psi \theta ^{\mu },\quad
\mathbf{S}=S_{\mu \nu }\theta ^{\mu \nu }=\frac{1}{2}\psi ^{\dagger }\gamma
_{0}i\gamma _{\mu \nu }\psi \theta ^{\mu }\wedge \theta ^{\nu }, \notag \\
\mathbf{K}& =K_{\mu }\theta ^{\mu }=\psi ^{\dagger }\gamma _{0}i\gamma
_{0123}\gamma _{\mu }\psi \theta ^{\mu },\quad \omega =-\psi ^{\dagger
}\gamma _{0}\gamma _{0123}\psi. \label{fierz}
\end{align}
 $\{\gamma _{\mu }\}$ denotes to the Dirac matrices, and the objects in (\ref{fierz}) satisfy the Fierz identities \cite{cra,lou2,holl}
 $\mathbf{J}^{2}=\omega^{2}+\sigma^{2},\;\;\;\mathbf{J}\llcorner\mathbf{K}=0,\;\;\;\mathbf{K}^{2}=-\mathbf{J}
^{2}$, and $\mathbf{J}\wedge\mathbf{K}
=-(\omega+\sigma\gamma_{0123})\mathbf{S}$. A spinor field such that at least one of the $\omega$ and the $\sigma$ are null [not null] is said to be singular [regular]. The Lounesto spinor field classification is provided by the following spinor field
classes \cite{lou2}:
\begin{itemize}
\item[1)] $\sigma\neq0,\;\;\; \omega\neq0$\qquad\qquad\qquad\qquad\qquad4) $\sigma= 0 = \omega, \;\;\;\mathbf{K}\neq 0, \;\;\;\mathbf{S}\neq0$%
\label{Elko11}
\item[2)] $\sigma\neq0,\;\;\; \omega= 0$\label{dirac1}\qquad\qquad\qquad\qquad\qquad5) $\sigma= 0 = \omega, \;\;\;\mathbf{K}=0,\;\;\; \mathbf{S}\neq0$%
\label{tipo41}
\item[3)] $\sigma= 0, \;\;\;\omega\neq0$\label{dirac21} \qquad\qquad\qquad\qquad\qquad\!6) $\sigma= 0 = \omega, \;\;\; \mathbf{K}\neq0, \;\;\; \mathbf{S} = 0$%
\end{itemize}
\noi
\noindent Types-(1), -(2), and -(3) are named {}{Dirac spinor fields} in the Lounesto classification, and in these cases it is implicit 
that $\mathbf{J}, \mathbf{K}, \mathbf{S}\neq0$. Types-(4), -(5), and -(6) are respectively called {}{flag-dipole}, {}{flagpole}, and {}{Weyl spinor fields}. For Dirac spinor fields, $\mathbf{S}$ is the distribution of intrinsic angular momentum; $\mathbf{J}$ is associated with the current of probability, and $\mathbf{K}$ is associated with the direction of the electron spin \cite{cra,lou2,holl}. 
By introducing the element $Z=\sigma+\mathbf{J}+i\mathbf{S} +i\mathbf{K}\gamma_{0123}+\omega\gamma_{0123} $, $Z$ is denominated a {boomerang} whenever it satisfies $\gamma_{0}Z^{\dagger}\gamma_{0}=Z$. When a spinor field is singular, namely it satisfies $\sigma = 0 = \omega$, the Fierz identities are substituted by the more general identities \cite{cra}:
\begin{align}
Z^{2} =\sigma Z,\quad Zi\gamma_{\mu\nu
}Z=4S_{\mu\nu}Z,\quad Z\gamma_{\mu}Z=4J_{\mu}Z, \quad
Zi\gamma_{0123}\gamma_{\mu}Z =4K_{\mu}Z,\quad Z\gamma_{0123}Z=-4\omega Z.
\end{align}
When one considers a type-(4) flag-dipole spinor field, 
the distribution of intrinsic angular momentum is provided by $\mathbf{S} = \mathbf{J}\wedge s$, where $s$ is a spacelike vector orthogonal to \textbf{J}. The real number $h\neq 0$ is such that \textbf{K} = $h$\textbf{J}, evincing thus the definition of helicity. It satisfies $h^2 = 1 + s^2$, implying the definition of helicity $h$ in quantum mechanics \cite{boehmergrafeno}. 
Type-(5) spinor fields are a particular case where $h=0$. Indeed, by $\mathbf{K} = h\mathbf{J}$, when $h=0$ the expressions $\omega=0=\sigma$, \textbf{K} $= 0$, \textbf{J} $\neq 0$ hold. Type-(5) spinor fields are therefore limiting cases of type-(4) spinor fields. 
More details on the most general form of type-(4) spinor fields are provided in the
Appendix.
\section{Matter Fields in Riemann-Cartan Geometries}
Once some features related to type-(4) spinor fields are introduced, we shall take into account the Wigner classification, to further study the spinor fields properties. According to the Wigner classification, in terms of irreducible representations of the Poincar\'{e} group, quantum particles are classified in terms of their mass and spin labels. The corresponding quantities for the quantum fields are given in terms of energy and spin densities. If one wishes to pursue the same spirit that Einstein followed to develop a theory for gravity, expressing the field equations by coupling the curvature to energy, in the most general case where torsion is present, one is compelled to recover the field equations coupling the curvature to energy but accompanied by similar field equations coupling the torsion to spin. When this is accomplished in the most straightforward way, the Einstein equations for the curvature-energy coupling are generalized as to include the Sciama-Kibble equations for the torsion-spin coupling. Namely, the ESK system of field equations, which can be obtained by generalizing the Ricci scalar written in terms of the metric $R(g)$ by the Ricci scalar written in terms of both metric and torsion $R(g,T)$ in the action, and subsequently varying it with respect to the two independent fields.

Notwithstanding, this is merely the most straightforward generalization of gravity with torsion. Other more general theories can be obtained by adding torsion not only implicitly through the curvature, but explicitly as well, as quadratic terms beside the curvature $R(g,T)+T^{2}$ in the action. Once the field equations are written down, and all torsional contributions are separated and evinced as spinor interactions, the effects of these extensions are reduced to a simple scaling of the torsional terms, or equivalently of the spinor interaction. It is evinced by introducing new coupling constants for such spin potentials. One of the most important problems about torsion in gravity, namely the fact that torsion should have been relevant only at the Planck scales, can thus be overcome since in these theories torsion has its own coupling constant, that does not necessarily coincide with the gravitational constant \cite{Fabbri:2011kq,Fabbri:2012yg}. 

On the other hand, however, those theories do not encompass the possibility to have \emph{dynamical} extensions, such as those provided by higher-order derivative field equations. The two most important ones are the case for which the Ricci scalar $R$ is replaced by an arbitrary function $f(R)$ in the action \cite{FV1}, and the one that is capable of implementing the conformal symmetry in the action itself \cite{Fabbri:2011ha,Fabbri:2011mi}. In the following we shall deal with both of them.
\subsection{Torsional $f(R)$-Gravity}

The extension of the Einstein-Hilbert action regarding an arbitrary function $f(R)$ is captivating, since it is the most general whenever one restricts the Ricci scalar as the sole source of dynamical information. In the case where both the metric and the torsion as well are taken into account, the variation with respect to an arbitrary metric $g$ and a $g$-compatible connection $\Gamma$ (or equivalently a tetrad field $e$ and a spin-connection $\omega$) yields the metric-affine (or tetrad-affine) approach(es) \cite{CCSV1,CCSV2,CV4,Rubilar}. The correspondent field equations are
\begin{subequations}
\label{2.1}
\begin{equation}
\label{2.1b }
\qquad\qquad\qquad T_{ij}^{\;\;\;h}
=\frac{1}{f'(R)}
\left[\frac{1}{2}\left(\de{f'(R)}/de{x^{p}}+S_{pq}^{\;\;\;q}\right)
\epsilon_{r}^{\;\,ph}\epsilon_{i\;\,j}^{\,r}
+S_{ij}^{\;\;\;h}\right],
\end{equation}
\begin{equation}
\label{2.1a}
\hspace*{-1.3cm}\Sigma_{ij}=f'\/(R)R_{ij} -\frac{1}{2}f\/(R)g_{ij},
\end{equation}
\end{subequations}
where $R_{ij}$, $\epsilon_{ijk}$, and $T_{ij}^{\;\;\;h}$ are the Ricci, the Levi-Civita, and the torsion tensors respectively. The $\Sigma_{ij}$ and $S_{ij}^{\;\;\;h}$ denote the stress-energy and spin density tensors associated to the matter fields: the conservation laws
\begin{subequations}
\label{2.2}
\begin{equation}
\label{2.2a}
\nabla_{i}\Sigma^{ij}+T_{i}\Sigma^{ij}-\Sigma_{pi}T^{jpi}-\frac{1}{2}S_{sti}R^{stij}=0,
\end{equation}
\begin{equation}
\label{2.2b}
\qquad\qquad\nabla_{h}S^{ijh}+T_{h}S^{ijh}+\Sigma^{ij}-\Sigma^{ji}=0,
\end{equation}
\end{subequations}
come from the Bianchi identities \cite{FV1}. In Eqs.\eqref{2.2} the symbols $\nabla_i$ and $R^{ijkl}$ denote respectively the covariant derivative and the curvature tensor, with respect to the dynamical connection $\Gamma$. By denoting $\Gamma^i = e^i_\mu\gamma^\mu$, where $e^\mu_i$ is a tetrad associated with the metric, and by introducing S$_{\mu\nu}:= \frac{1}{8}[\gamma_\mu,\gamma_\nu]$, the covariant derivatives of the matter field $\psi$ and its Dirac adjoint are denoted by $D_i\psi = \de\psi/de{x^i} + \omega_i^{\;\;\mu\nu}{\rm S}_{\mu\nu}\psi\/$ and $D_i\bar\psi = \de{\bar\psi}/de{x^i} - \bar\psi\omega_i^{\;\;\mu\nu}{\rm S}_{\mu\nu}\/$, where $\omega_i^{\;\;\mu\nu}$ is the spin connection. One can furthermore indite $D_i\psi = \de\psi/de{x^i} - \Omega_i\psi$ and $D_i\bar\psi = \de{\bar\psi}/de{x^i} + \bar{\psi}\Omega_i$ where
\begin{equation}\label{2.3}
\Omega_i := - \frac{1}{4}g_{jh}\left(\Gamma_{ik}^{\;\;\;j} - e^j_\mu\partial_i\/e^\mu_k \right)\Gamma^h\Gamma^k.
\end{equation}
$\Gamma_{ik}^{\;\;\;j}$ denote the coefficients of the linear connection $\Gamma$, since the relation between linear and spin connection is provided by
$\Gamma_{ij}^{\;\;\;h} = \omega_{i\;\;\;\nu}^{\;\;\mu}e_\mu^h\/e^\nu_j + e^{h}_{\mu}\partial_{i}e^{\mu}_{j}
$, 
as can be immediately calculated.
In the case of matter fields, the spin density tensor is given by S$_{ij}^{\;\;\;h}=\frac{i}{2}\bar\psi\left\{\Gamma^{h}, {\rm S}_{ij}\right\}\psi \equiv-\frac{1}{4}\eta^{\mu\sigma}\epsilon_{\sigma\nu\lambda\tau} K^\tau e^{h}_{\mu}e^{\nu}_{i}e^{\lambda}_{j}$. Remember that $K^\tau$ is the component of the pseudo-vector bilinear covariant defined at (\ref{fierz}). The stress-energy tensor components of the matter fields are hence described as
\begin{equation}\label{2.5}
\Sigma^D_{ij} := \frac{i}{4}\/\left( \bar\psi\Gamma_{i}{D}_{j}\psi - {D}_{j}\bar{\psi}\Gamma_{i}\psi \right)\quad\quad\text{and}\quad\;\;\;
\Sigma^F_{ij}:= (\rho +p)\/U_iU_j -pg_{ij}.
\end{equation}
In Eqs.\eqref{2.5}, $\rho$, $p$ and $U_i$ denote respectively the matter-energy density, the pressure, and the four-velocity of the fluid. The trace of the equations \eqref{2.1a}, given by \begin{equation}\label{2.6}
f'(R)R -2f(R)=\Sigma,
\end{equation} is supposed
to relate the Ricci scalar curvature $R$ and the trace $\Sigma$ of the stress-energy tensor, as in \cite{CCSV1,CCSV2,CV4,FV1}. Furthermore, it is assumed that $f(R)\not = kR^2$ --- since the case $f(R)=kR^2$ is solely compatible to the condition $\Sigma=0$. Now, from Eq.\eqref{2.6} it is possible to express $R=F(\Sigma)$, where $F$ is an arbitrary function. Furthermore, introducing the scalar field
$\varphi := f'\/(F\/(\Sigma))$ as well as the effective potential $V(\varphi):= \frac{1}{4}\left[ \varphi F^{-1}\/((f')^{-1}\/(\varphi))+ \varphi^2\/(f')^{-1}\/(\varphi)\right]$, the field equations \eqref{2.1a} are written in the Einstein-like form
\begin{equation}\label{2.9}
\begin{split}
\mathring{R}_{ij} -\frac{1}{2}\mathring{R}g_{ij}= \frac{1}{\varphi}\Sigma^F_{ij} + \frac{1}{\varphi}\Sigma^D_{ij} + \frac{1}{\varphi^2}\left( - \frac{3}{2}\varphi_i\varphi_j + \varphi\mathring{\nabla}_{j}\varphi_i + \frac{3}{4}\varphi_h\varphi_k g^{hk}g_{ij} \right. \\
\left. - \varphi\mathring{\nabla}^h\varphi_hg_{ij} - V\/(\varphi)g_{ij} \right) + \mathring{\nabla}_h\hat{{\rm S}}_{ji}^{\;\;\;h} + \hat{{\rm S}}_{hi}^{\;\;\;p}\hat{{\rm S}}_{jp}^{\;\;\;h} - \frac{1}{2}\hat{{\rm S}}_{hq}^{\;\;\;p}\hat{{\rm S}}_{\;\;p}^{q\;\;\;h}g_{ij},
\end{split}
\end{equation}
where $\mathring{R}_{ij}$, $\mathring R$ and $\mathring{\nabla}_i$ denote respectively the Ricci tensor, the Ricci scalar curvature and the covariant derivative of the Levi--Civita connection. Here $\hat{{\rm S}}_{ij}^{\;\;\;h}:=-\frac{1}{2\varphi}{\rm S}_{ij}^{\;\;\;h}\/$ and $\varphi_i:=\frac{\partial\varphi}{\partial x^i}$.
In addition, the generalized Dirac equations for the spinor field are in this context
\begin{equation}\label{2.10}
i\Gamma^{h}D_{h}\psi + \frac{i}{2}T_h\Gamma^h\psi- m\psi=0,
\end{equation}
where $T_h :=T_{hj}^{\;\;\;j}$ is the axial torsion\footnote{It is interesting to note that at this point it is not formally explicit by \eqref{2.10} whether we are dealing to Dirac equation with torsion in a simply connected space or with a Dirac equation without torsion in a multiply connected space-time \cite{Roc11}. As both descriptions are mathematically equivalent, we can transpose one formalism into another, in order to circumvent such
question.}. The symmetrized part of the Einstein-like equations \eqref{2.9} as well as the Dirac equations \eqref{2.10} are written as \cite{FV1}
\begin{equation}\label{2.15}
\begin{split}
\mathring{R}_{ij} -\frac{1}{2}\mathring{R}g_{ij}= \frac{1}{\varphi}\Sigma^F_{ij} + \frac{1}{\varphi}\mathring{\Sigma}^D_{ij}
+ \frac{1}{\varphi^2}\left( - \frac{3}{2}\varphi_i\varphi_j + \varphi\mathring{\nabla}_{j}\varphi_i +
\frac{3}{4}\varphi_h\varphi_kg^{hk}g_{ij} \right. \\
\left. - \varphi\mathring{\nabla}^h\varphi_hg_{ij} - V\/(\varphi)g_{ij} \right) + \frac{3}{64\varphi^2}K^\tau K_\tau g_{ij}
\end{split}
\end{equation}
and
\begin{equation}\label{2.16}
i\Gamma^{h}\mathring{D}_{h}\psi
-\frac{3}{16\varphi}\left[\sigma
+i\omega\gamma_5\right]\psi-m\psi=0,
\end{equation}
where
$\mathring{\Sigma}^D_{ij} := \frac{i}{4}\/\left[ \bar\psi\Gamma_{(i}\mathring{D}_{j)}\psi - \left(\mathring{D}_{(j}\bar\psi\right)\Gamma_{i)}\psi \right]$ and $\mathring{D}_i$ is the covariant derivative of the Levi--Civita connection.

As spinor fields satisfying the Dirac equation in this scenario are incompatible with stationary spherical symmetry \cite{Fabbri:2011mg}, the simplest choice for the background must be at least an axially symmetric Bianchi-I type metric, given by the form $ds^2 = dt^2 - a^2(t)\,dx^2 - b^2(t)\,dy^2 - c^2(t)\,dz^2$, where the $\Gamma^i = e^i_\mu\gamma^\mu$ are given by
\begin{equation}\label{3.5}
\Gamma^0 = \gamma^0,\qquad \Gamma^1 = \frac{1}{a(t)}\gamma^1, \qquad \Gamma^2 = \frac{1}{b(t)}\gamma^2, \qquad \Gamma^3 = \frac{1}{c(t)}\gamma^3,
\end{equation} and the tetrad field is given by $e^\mu_0=\delta^\mu_0$, $e^\mu_1 = a(t)\/\delta^\mu_1$, $e^\mu_2 = b(t)\/\delta^\mu_2,$ and $e^\mu_3 = c(t)\/\delta^\mu_3$, for $ \mu =0,1,2,3$. The spin-Dirac operator acts on spinor fields and their conjugates respectively as $\mathring{D}_i\psi = \partial_i\psi - \mathring{\Omega}_i\psi$ and $ \mathring{D}_i\bar\psi = \partial_i\bar\psi + \bar{\psi}\mathring{\Omega}_i$, where the spin connection coefficients $\mathring{\Omega}_i$ are given by (introducing the notation $a_1 = a$, $a_2 = b$, and $a_3 = c$) \[\mathring{\Omega}_0=0,\qquad\qquad\quad \mathring{\Omega}_i=\frac{1}{2}{\dot{a}_i}\gamma^i\gamma^0.\]

Therefore, the Einstein-like equation \eqref{2.15} reads\begin{subequations}\label{3.10}
\begin{equation}\label{3.10a}
\begin{split}
\frac{\dot a}{a}\frac{\dot b}{b} + \frac{\dot b}{b}\frac{\dot c}{c} + \frac{\dot a}{a}\frac{\dot c}{c} = \frac{\rho}{\varphi} - \frac{3}{64\varphi^2}K^\sigma\,K_\sigma\,
+\frac{1}{\varphi^2}\left[- \frac{3}{4}{\dot\varphi}^2 - \varphi\dot\varphi\frac{\dot\tau}{\tau} - V(\varphi)\right],
\end{split}
\end{equation}
\begin{equation}\label{3.10b}
\begin{split}
\frac{\ddot a_r}{a_r} + \frac{\ddot a_s}{a_s} + \frac{\dot a_r}{a_r}\frac{\dot a_s}{a_s} = - \frac{p}{\varphi} +
\frac{1}{\varphi^2}\left[\varphi\dot\varphi\frac{\dot a_t}{a_t}+\frac{3}{4}{\dot\varphi}^2 -\varphi\left( \ddot\varphi + \frac{\dot\tau}{\tau}\dot\varphi \right) - V(\varphi)\right] +\frac{3}{64\varphi^2}K^\sigma\,K_\sigma\, ,
\end{split}
\end{equation}
\end{subequations} where $r,s,t$ denote indexes $1,2,3$ different from each other. The Dirac field equation \eqref{2.16} assumes the form
\begin{equation}\label{3.9}
\dot\psi + \frac{\dot\tau}{2\tau}\psi + im\gamma^0\psi - \frac{3i}{16\phi}(\sigma\gamma^0 + i\omega\gamma^0\gamma^5) \psi = 0,
\end{equation}
where $\tau := abc$ \cite{Saha1,Saha2}. Together with the conditions
\begin{equation}\label{3.11}
\mathring{\Sigma}^D_{rs}=0\quad \Rightarrow \quad a_r\/\dot{a}_s - a_s\/\dot{a}_r=0 \quad \cup \quad K^\intercal =0,
\end{equation}
the equations $\mathring{\Sigma}^D_{0i}=0$ are automatically satisfied. Finally, the conservation laws together with an equation of state of the kind $p=\lambda\rho$ (here $\lambda$ is a number between 0 and 1) yield $\dot\rho + \frac{\dot\tau}{\tau}(1+\lambda)\rho =0$, which completes the whole set of field equations, having the general solution given by
\begin{equation}\label{3.12bis}
\rho = \rho_0\tau^{-(1+\lambda)}\,, \qquad \rho_0 = {\rm constant}.
\end{equation}

The matter field in such axially symmetric background is such that conditions \eqref{3.11} are constraints imposed on the metric or on the matter field. They exist if and only if one of the following conditions holds:
\begin{enumerate}
\item by imposing constraints of purely geometrical origin, as $a\dot{b}-b\dot{a}=0$, $a\dot{c}-c\dot{a}=0$, $c\dot{b}-b\dot{c}=0$. In this scenario there are fermionic matter fields in an isotropic Universe, which might \emph{a priori} cause some pathology, as Dirac fields are well known not to undergo the Cosmological Principle \cite{t}. But the result by Tsamparlis \cite{t}, although valid for Dirac spinor fields, does not hold for the other spinor field classes, according to Lounesto classification.

\item another condition is to impose constraints of purely material origin by requiring that the spatial components of the spin direction satisfy $K_i = 0$. This represents an anisotropic Universe devoid of terms coupling matter to the axial torsion. In this case there is no fermionic torsional interactions. Indeed, the particle spin interacts with the axial component of the torsion tensor, and when the spatial components of the spin direction equal zero it implies that such particles described by the field $\psi$ do not interact to the torsion. Besides, if Dirac fields are absent then it is not clear what may then justify anisotropies \cite{fab}.

\item the last situation would be originated by the geometry and the matter as well, by insisting that for instance $a\dot{b}-b\dot{a}=0$ and $K_1 = 0 = K_2$. It provides partial isotropy for only two axes, with the corresponding components of the spin vector vanishing.
It describes a Universe an ellipsoid of rotation about the axis along which the spin vector does not vanish. By insisting on the proportionality between two pairs of axes we inevitably get the total isotropy of the $3$-dimensional space. Therefore, the situation in which we have $a=b$, with $K_1 = 0 = K_2$, is the only one be entirely satisfactory. Henceforth this situation shall be considered, where the sole spatial component of the spin direction is $K_3\neq 0$.
\end{enumerate}
Here, the Dirac and Einstein-like equations \eqref{3.10} and \eqref{3.9} can be worked out as in \cite{Saha1,Saha2}: for instance, through suitable combinations of \eqref{3.10} we obtain the equations
\begin{subequations}\label{3.13}
\begin{equation}\label{3.13a}
\frac{d}{dt}(J_0\tau)=0 =
\frac{d}{dt}(\sigma\tau)
+\frac{3\omega K_0\tau}{8\varphi},
\end{equation}
\begin{equation}\label{3.13c}
-\frac{d}{dt}(\omega\tau)
+\left[2m+\frac{3\sigma}{8\varphi}\right]K_0\tau=0=\frac{d}{dt}(K_0\tau)+2m\omega\tau.
\end{equation}
\end{subequations}
while from Eqs.\eqref{3.13} it is straightforward to deduce that
\begin{equation}\label{3.13abis}
(K_3)^2=\sigma^2 + \omega^2 + (K_0)^2 = \frac{C^2}{\tau^2},\qquad\qquad
(J_0)^2=\frac{D^2}{\tau^2},
\end{equation}
with $C$ and $D$ constants. It is worthwhile to emphasize that in this special case the theory has an additional discrete symmetry provided by the transformation $\psi \mapsto \gamma^5\gamma^0\gamma^1\psi$, making all field equations are invariant. In the Dirac equation the four complex components is in this case reduced to two complex components. Such assertion is equivalent to take flagpole spinor fields, that have four real parameters. Hence \eqref{3.13} are the field equations to be solved. The compatibility to all constraints allows only three classes of spinor fields, each of which has a general member written in one of the following form
\begin{eqnarray}
\nonumber
\label{generalspinor}
&\psi=\frac{1}{\sqrt{2\tau}}\left(\begin{tabular}{c}
$\sqrt{K-C}\cos{\zeta_{1}}e^{i\theta_{1}}$\\
$\sqrt{K+C}\cos{\zeta_{2}}e^{i\vartheta_{1}}$\\
$\sqrt{K-C}\sin{\zeta_{1}}e^{i\vartheta_{2}}$\\
$\sqrt{K+C}\sin{\zeta_{2}}e^{i\theta_{2}}$
\end{tabular}\right),
\end{eqnarray}
with constraints $\tan{\zeta_{1}}\tan{\zeta_{2}}=(-1)^{n+1}$ and $\theta_{1}+\theta_{2}-\vartheta_{1}-\vartheta_{2}=\pi n$ for any $n$ integer, and also
\begin{eqnarray}
\label{restrictedspinor1}
&\psi=\frac{1}{\sqrt{2\tau}}\left(\begin{tabular}{c}
$\sqrt{K-C}\cos{\zeta_{1}}e^{i\theta_{1}}$\\
$0$\\
$0$\\
$\sqrt{K+C}\sin{\zeta_{2}}e^{i\theta_{2}}$
\end{tabular}\right)\quad
\text{and}\;\;\;
\psi=\frac{1}{\sqrt{2\tau}}\left(\begin{tabular}{c}
$0$\\
$\sqrt{K+C}\cos{\zeta_{1}}e^{i\vartheta_{1}}$\\
$\sqrt{K-C}\sin{\zeta_{2}}e^{i\vartheta_{2}}$\\
$0$
\end{tabular}\right).
\end{eqnarray}
where $\zeta_{1}$, $\zeta_{2}$, $\theta_{1}$, $\theta_{2}$, $\vartheta_{1}$, $\vartheta_{2}$ are time dependent. The most interesting case is one the provided by \eqref{restrictedspinor1}. For instance the second spinor field at \eqref{restrictedspinor1} is
\begin{eqnarray}
\label{spinorsolution}
&\psi=\frac{1}{\sqrt{2\tau}}\left(\begin{tabular}{c}
$0$\\
$\sqrt{K+C}e^{i\beta(t)}$\\
$\sqrt{K-C}e^{-i\beta(t)}$\\
$0$
\end{tabular}\right),\quad\text{}
\end{eqnarray} for $\beta(t)=-mt-\frac{3C}{16}\int{\frac{dt}{\tau}}$. There are further constraints $\sigma=\frac{C}{\tau}$, $\psi^{\dagger}\psi=\frac{K}{\tau}$ and $\omega=0 = K_0$. Such a spinor field {\it is a type-(4) flag-dipole spinor field, according to the Lounesto spinor fields classification} \cite{plbb}. This is a remarkable fact: once it is assumed a spinor field $\psi$ in a $f(R)$ Riemann-Cartan cosmology, some type-(4) spinor fields are obtained as the spinor fields (\ref{restrictedspinor1}). Indeed, there is no assumption in Eq.(\ref{2.10}) that makes $\psi$ a legitimate Dirac spinor field, as it merely regards \emph{a priori} a spinor field $\psi$ that satisfies the Dirac equation. As far as we know, this is up to now the unique physical system whose acceptable solution is given in terms of such spinor fields.

On the other hand, when one imposes $K_3 = 0$ as a constraint of purely material origin, Eqs.(\ref{3.13abis}) implies that $K_0 = 0$. Therefore $K^\mu = 0$ and we obtain a type-(5) spinor field under Lounesto spinor field classification, which encompasses Majorana and Elko dark spinor fields. It must be stressed that the condition $K_3=0$ does not necessarily imply that in this case there is no fermion fields satisfying the Dirac equation (\ref{2.10}). In fact Elko fields do not satisfy the Dirac equation at all\footnote{In fact, Elko spinor fields \cite{allu,allu1} are eigenspinors of the charge conjugation operator and do not satisfy Dirac equation \cite{allu,allu1}. Some important applications are provided, for instance, at \cite{Basak:2012sn}. There is still the complementary set of Elko and Majorana fields, with respect to the type-(5) spinor, whose dynamics is still unknown. Its general form is provided in the Appendix.}.

In summary, by the solutions above, the so-called Dirac field $\psi$ in (\ref{restrictedspinor1}, \ref{spinorsolution}) is not a Dirac spinor field according to Lounesto classification, but a type-(4) flag-dipole spinor field. Besides, since $K_i = 0$ and in particular $K_3=0$, by \eqref{3.13abis} it implies that we are concerning now a type-(5) spinor field, which is a flagpole. But in this case, it is well-known that type-(5) encompasses Elko, Majorana, and the complementary spinor fields, presented at \eqref{eq10}. Elko, however, is well known not to satisfy Dirac equations, so as we departure from \eqref{3.13abis}, Elko is excluded to be a solution of such system. The point to be stressed here is that according to the Lounesto spinor field classification, $\psi$ can be allocated in any of the six disjoint classes and there is no \emph{ab initio} relation between the type of the spinor field and the associated dynamics. As mentioned, for instance, the types-(1), (2) and (3) are Dirac spinor fields in the Lounesto classification, having some subset satisfying the Dirac equation. By the same token, type-(6) spinor fields encompass Weyl spinor fields, that indeed satisfy Dirac equations. Nevertheless, it was an open problem whether type-(4) spinor fields satisfy or not the Dirac equations, but the Dirac equations is shown to be dynamically forbidden for the solutions found \cite{fab}.
\subsection{Torsional Conformal-Gravity}

It is worth to point some recent progress in the study of spinor fields in generalized gravity, as well as some open issues which are under current investigation. While it is somewhat apart of the main theme of the paper, it is certainly enriching from the bookkeeping purposes. In this vein, another interesting higher-order theory of gravity is the one with two curvatures, because this is the only case in which conformal invariance can be obtained \cite{Fabbri:2011ha}. As it turns out, there are two ways to implement conformal transformations for torsion: the first is to require the most general (reasonable) conformal transformation for torsion (where by reasonable we mean reasonable according to what is discussed, for instance, in \cite{sh}). The another is to insist on the fact that no conformal transformation is to be given to torsion (because conformal transformations are of metric origin while torsion is independent on the metric). In the former case, because conformal transformations link the metric to torsion, one must modify the Riemann curvature with quadratic-trace torsion terms in order to get a curvature whose irreducible part is conformally invariant \cite{Fabbri:2011ha}. In the latter case, torsion and curvature are separated and essentially independent. Consequently, in the former case \cite{Fabbri:2011ha} the field equations are closely intertwined together, while in the latter case the field equations are independent thus maintaining the curvature-energy and torsion-spin coupling in the spirit of the ESK field equations.

\subsubsection{Torsion with general Conformal Transformations}
In the first case the coupling to the Dirac field has been studied in \cite{Fabbri:2011ha}. However, as in this case the field equations that couple torsion to spin are not invertible in general, torsion cannot be substituted by the spin density into the Dirac field equations, which therefore remain of the general form
\begin{eqnarray}
&i\gamma^{\mu}\mathring{D}_{\mu}\psi+\frac{3}{4}W_{\sigma}\gamma^{5}\gamma^{\sigma}\psi=0,
\end{eqnarray}
where $W_{\sigma}$ is the axial vector dual of the completely antisymmetric part of the torsion tensor. Hence the arguments used in \cite{Fabbri:2011mg} cannot be recovered, and therefore stationary spherically symmetric symmetries are possible. However in such a case, the complete antisymmetry of the Dirac field does not turn into the complete antisymmetry of torsion. Instead, rather in constraints for the gravitational fields that cannot be satisfied in general situations. In this case of general torsional conformal transformations the Dirac field appears to be ill-defined.

An alternative situation is therefore to study Elko fields, which has been accomplished in \cite{Fabbri:2011mi}. However, their dynamics in terms of cosmological solutions has not been studied yet.

\subsubsection{Torsion with no Conformal Transformations}
The coupling to the Dirac field was studied \cite{Fabbri:2011ha}, showing that the complete antisymmetry of the spin density results into the complete antisymmetry of the torsion tensor, which dual is an axial vector given by
\begin{eqnarray}
&W_{\rho}=\left(\frac{4a}{\hbar}\,K^{\mu}K_{\mu}\right)^{-1/3}J_{\rho},
\label{torsionequations}
\end{eqnarray}
so that torsion can be replaced with the spin density of the spinor field, and the Dirac field equation becomes
\begin{eqnarray}
&i\gamma^{\mu}\mathring{D}_{\mu}\psi
-\left(\frac{256a}{27}K^{\rho}K_{\rho}\right)^{-\frac{1}{3}} \overline{\psi}\gamma_{\nu}\psi\gamma^{\nu}\psi=0,
\label{matterequations}
\end{eqnarray}
with a non-linear self-interaction that is renormalizable nonetheless. After a straightforward Fierz rearrangement they can be written as
\begin{eqnarray}
&i\gamma^{\mu}\mathring{D}_{\mu}\psi
-\left(\frac{27}{256a}\right)^{{1}/{3}}
{}{}\left(\sigma^{2}
{}{}+{}i\omega^2\right)^{-{1}/{3}}
{}\left(\sigma \mathbb{I}{}
{}-\omega\gamma_5\right)\psi=0,
\label{matteqarranged}
\end{eqnarray}
clearly showing that the type-(4) spinor fields would verify a Dirac field equation of the form
 $i\gamma^{\mu}\mathring{D}_{\mu}\psi=0$, as if torsion were never present, precisely like the ESK theory. In this case, it again happens that the reasoning performed in \cite{Fabbri:2011mg} does not apply, and stationary spherically symmetric solutions are possible, the gravitational field equations would reduce to the torsionless spherically symmetric Weyl field equations in a Schwarzschild spacetime.
For this type of conformal gravity, the case of Elko fields has not been studied yet.
\section{Conclusions}
In this paper, we have explored both the regular and singular spinor fields, establishing the general gravitational background with torsion in which the spinor fields are supposed to live in.
We proved that some singular flag-dipoles spinor fields are physical solutions for the Dirac equation in ESK theories: in particular this has been obtained in $f(R)$-gravity but it could not be recovered in conformal gravity as well.

In the case of cosmology, when considering Dirac-like fields in $f(R)$-gravity, the presence of torsion imposes the use of an anisotropic background in which the geometric side is diagonal, while the energy tensor is not, due to intrinsic features of the spinor field. In this circumstance, the non-diagonal part of the gravitational field equations results into the constraints \eqref{3.11} characterizing the structure of the spacetime, or the helicity of the spinor field, or both. In our understanding, the only physically meaningful situation is the one in which two axes are equal and one spatial component of the axial vector torsion does not vanish. It provides a Universe that is spatially an ellipsoid of rotation revolving about the only axis along which the spin density is not equal to zero. 

In the case of conformal gravity, except for the case of torsional conformal transformations, for which the Dirac field seems not well-defined, the case of torsion without conformal transformations appears to be well-posed. In this case, the gravitational background is much like the torsionless one, and although we have not proved it mathematically, there are reasons to believe that singular type-(4) spinor fields may still emerge.

In summary, the presence of torsion induces non-linear interactions, whose details depend on whatever conformal of $f(R)$ gravitational background is used, but in general such torsionally-induced self-interactions for the spinors affect the dynamics of the spinor itself: specifically, it is possible to find perfectly physical solutions of the Dirac equation which are nevertheless not Dirac fields, but flag-dipoles, and thus singular. We have also found that, in addition, the new solutions encompass Elko and Majorana spinor fields, when the associated spin direction vanishes, providing an anisotropic Universe without fermionic torsional interactions.

However, we believe that the main message that is to be taken is that a spinor field satisfying the Dirac field equation is not necessarily non-singular: with a metaphoric analogy, we may say that the Dirac equations does not necessarily take care of itself by forbidding singular solutions. 

To remove them, an even deeper analysis must be carried over at an algebraic level.

\section*{Acknowledgments}

R. da Rocha and is grateful to CNPq grants 451042/2012-3 for financial support. R. T. Cavalcanti is grateful to CAPES and to UFABC for financial support. J. A. Silva-Neto is grateful to UFABC for financial support. J. M. Hoff da Silva in grateful to CNPq for financial support.

\appendix

\section{Clifford Algebras and General type-(4) and type-(5) spinor fields}
\label{appa}
Let $V$ be a finite $n$-dimensional real vector space and $\Lambda(V) = \bigoplus_{k=0}^n\Lambda^k(V)$ the space of multivectors over $V$, where $\Lambda^k(V)$ denotes the $k$-forms vector space. By defining the reversion, given $\tau,\psi,\xi\in\Lambda(V)$, the { left contraction} is defined implicitly by $\eta(\tau\lrcorner\psi,\xi)=\eta(\psi,\tilde\tau\w\xi)$. The Clifford product between $\vv\in V$ and $\psi$ is provided by $\vv\psi = \vv\w \psi + \vv\lrcorner \psi$. Given a metric $\eta$, the pair $(\la(V),\eta)$ endowed of the Clifford product is the Clifford algebra $\cl_{1,3}$ of $\RR^{1,3}$. All spinor fields are placed in a manifold which locally is a Minkowski spacetime $(M,\eta,\mathring{D},\tau_{\eta },\uparrow )$ in what follows, where $M$ is a manifold, $\mathring{D}$ denotes the Levi-Civita connection associated with $\eta $, $M$ is oriented by the 4-volume element $\tau _{\eta }$ and time-oriented by $\uparrow $. Furthermore, $\{\mathbf{e}_{\mu }\}$ is a section of the frame bundle $\mathbf{P}_{\mathrm{SO}_{1,3}^{e}}(M)$. $\{\mathbf{e}^{\mu }\}$ is the dual frame: $\mathbf{e}^{\mu }(\mathbf{e}_{\nu})=\delta _{\nu }^{\mu }$, with $\{\theta ^{\mu }\}$ and $\{\theta _{\mu }\}$ respectively the dual bases \ of \ $\{\mathbf{e}_{\mu }\}$ and $\{\mathbf{e}^{\mu }\}$. Hereupon we denote $\mathbf{e}_{\mu\nu} = \mathbf{e}_{\mu}\mathbf{e}_{\nu}$ and $\mathbf{e}_{\mu\nu\rho} = \mathbf{e}_{\mu}\mathbf{e}_{\nu}\mathbf{e}_{\rho}$.

In order to better understand the structure of type-(4) and their limiting case type-(5) spinor fields, the question is: what is the general form of these spinor fields? In order to answer it, let us take a general spinor given by $\psi=(f, g, \eta, \xi)^\intercal $, with $ f,g,\eta,\xi \in \mathbb{C}$, and the definition of these spinor types given by Lounesto classification \cite{lou2}.
\subsection*{Spinor Fields of Type-(4)}
As we aim to characterize the most general type-(4) flag-dipole spinor field, the conditions $\sigma = 0 =\omega$ results $\eta f^{*}+\xi g^{*}=0$. We have to analyze the possibilities evinced from this equation. If $f=0=g$ or $\eta=0=\xi$, it implies a type-(6) spinor field, with $\textbf{S}=0$, and therefore this possibility must be dismissed here. It remains the conditions: either $\eta=0=\xi,\; f=0=g$, or none of the components can be zero. In this last case, one can isolate a part of them, for example $f=\frac{g\eta \xi^*}{\Vert \eta \Vert^2}$. Further, the condition $\textbf{K}\neq 0$ induces the following possibilities:
\begin{enumerate}
\item If $\eta=0=\xi$, hence $K_1 = K_2=0$, and $K_0 \neq 0 \neq K_3 \Rightarrow \Vert f \Vert^2 \neq \Vert \xi \Vert^2$;
\item If $f=0=\xi$, it implies that $K_1 = K_2=0$, and $K_0 \neq 0 \neq K_3 \Rightarrow \Vert g\Vert^2 \neq \Vert \eta\Vert^2$;
\item If all the components are not zero, $K_1 \neq 0 \neq K_2 \Rightarrow \Vert g\Vert^2 \neq \Vert \eta\Vert^2$.
\end{enumerate}
In the third case, if $\Vert g\Vert^2 = \Vert \eta\Vert^2$, therefore $\textbf{K}=0$. Furthermore, still in the third case, $\Vert g\Vert^2 \neq \Vert \eta\Vert^2 \Leftrightarrow \Vert f\Vert^2 \neq \Vert \xi\Vert^2$. Thus, the possible type-(4) spinor fields are:
\begin{eqnarray}\label{errr}\psi_{_{(4)}}&=&(f, 0, 0, \xi)^\intercal\;,\quad\qquad\quad\qquad \; \Vert f\Vert^2 \neq \Vert \xi\Vert^2\;\,,\;\;\;\;\;\;\;\;\;\text{or}\nonumber\\
\psi_{_{(4)}}&=&(0, g, \eta, 0)^\intercal\;,\quad\qquad\quad\qquad \;\Vert g\Vert^2 \neq \Vert \eta\Vert^2\;\,,\;\;\;\;\;\;\;\;\;\text{or}\nonumber\\
\psi_{_{(4)}}&=&\left(\frac{g\eta \xi^*}{\Vert \eta\Vert^2}\;, g, \eta, \xi\right)^\intercal,\qquad\quad\;\;\; \Vert g\Vert^2 \neq \Vert \eta \Vert^2 \;.\end{eqnarray}
If some inequality associated to one of the spinors above does not hold, it turns forthwith to be a type-(5), which shall be analyzed in what follows.

\subsection*{Spinor Fields of Type-(5)}
We start by noticing how the conditions on the bilinear covariants associated to a type-(5) spinor field
imply the following conditions on the spinor field components:
\begin{eqnarray}\label{eq1}
\sigma&=&\overline{\psi} \psi = 0 = -\overline{\psi}\gamma_{0123} \psi = \omega \Rightarrow \eta f^{*}+\xi g^{*}=0,\\\label{eq2}
K_1&=&\overline{\psi}i \gamma_{0123} \gamma_1 \psi = 0 = \overline{\psi}i \gamma_{0123} \gamma_2 \psi = K_2 \Rightarrow g f^{*}+\xi \eta^{*}=0,\\
\label{eq3}
K_0=\overline{\psi}i \gamma_{0123} \gamma_0 \psi &=& 0 = \overline{\psi}i \gamma_{0123} \gamma_3 \psi = K_3 \Rightarrow \|f\|^2=\|\xi\|^2 \mbox{ and } \|g\|^2=\|\eta\|^2.
\end{eqnarray}
Eq. (\ref{eq3}) can be obtained from (\ref{eq1}) and (\ref{eq2}), which are therefore essential to characterize type-($5$) spinor fields. In this vein, an equation candidate to describe the dynamics of these general spinor fields must keep \eqref{eq1} and \eqref{eq2} invariant. Elko spinor fields obey these equations.

By performing a straightforward calculation with the aid of Eqs.(\ref{eq1}) and (\ref{eq2}) it is possible to obtain
\begin{equation}\label{eq4}
f=-\xi^*(\eta + g)(\eta^*+g^*)^{-1}=-\xi^*\left(\frac{\eta+g}{\Vert \eta+g \Vert}\right)^2,
\end{equation}
and by taking $\tan\varphi_1=-i\frac{\eta+g-(\eta+g)^*}{\eta+g+(\eta+g)^*}$, we can write
$
f=-\xi^*e^{2i \varphi_1}$ and $g=-\eta^*e^{2i \varphi_2},
$
where $\varphi_1$ and $\varphi_2$ are related by \footnote{When $\varphi_1 \neq n \pi$, that is, $\eta$ + g is not real.}
\begin{equation}\label{eq7}
\tan\varphi_2=-i\frac{\xi(1+e^{-2i\varphi_1})-[\xi(1+e^{-2i\varphi_1})]^*}{\xi(1-e^{-2i\varphi_1})+[\xi(1-e^{-2i\varphi_1})]^*}=-\cot \varphi_1.\nonumber
\end{equation}
However, $\tan\varphi_2= -\cot \varphi_1 \Rightarrow \varphi_2= \varphi_1+(2k+1)\frac{\pi}{2}$, and then $e^{2i \varphi_2}=e^{2i \varphi_1}e^{i(2k+1)\pi}=-e^{2i \varphi_1}$, for every $k=0,1,2,\ldots$. Hence a general type-($5$) spinor can be represented by
\begin{eqnarray}
&\psi_{_{(5)}}=\left(-\xi^*e^{2i \varphi_1}, \eta^*e^{2i \varphi_1}, \,\eta, \,\xi\right)^\intercal\,.\label{eq8}
\end{eqnarray}
Writing $\psi_{_{(5)}} = (\chi_2, \chi_1)^\intercal$, it is straightforward to realize that $\chi_2= -i\sigma_2\chi_1^* e^{2i\varphi_1}=\sigma_2\chi_1^* e^{i(2\varphi_1-\frac{\pi}{2})}$. By taking $\varphi\equiv 2\varphi_1-\frac{\pi}{2}$ a more compact form of (\ref{eq8}) is
\begin{eqnarray}
\psi_{_{(5)}}=\left(e^{i\varphi}\sigma_2\chi_1^*\,,\,\chi_1\right)^\intercal\,.\label{eq10}
\end{eqnarray}
By acting now the charge conjugation operator \cite{allu,allu1}, with $i\Theta = \sigma_2$, it yields
\begin{equation}
C\psi_{_{(5)}}=\mu \psi_{_{(5)}},\qquad{\rm for}\quad C=%
\bigl(\begin{smallmatrix}
\OO & i\Theta \\
-i\Theta & \OO \nonumber
\end{smallmatrix}\bigr) {\cal K}
\qquad {\rm and}\quad \mu =-e^{i\varphi}
 .\label{conj}
\end{equation} Here ${\cal K}$ conjugates the spinor components. Hence the eigenvalues take place on the sphere $S^1$. When these eigenvalues are real and $\chi_1, \chi_2$ are dual helicity eigenstates, Elko spinor fields are obtained. The type-(5) flagpole spinor fields were shown 
to have a prominent role on the derivation of all Lagrangians for the gravity from the one for supergravity 
\cite{where1,where2}.

\appendix

\begin{thebibliography}{99}
\bibitem{h-h-k-n}
F.~W.~Hehl, P.~Von Der Heyde, G.~D.~Kerlick, J.~M.~Nester, {Rev. Mod. Phys.} \textbf{48}, 393 (1976).

\bibitem{v-z}
G.~Velo and D.~Zwanziger,
{Phys. Rev.} \textbf{186}, 1337 (1969).

\bibitem{lou2} P. Lounesto, {}{\it ``Clifford Algebras and Spinors''},
2$^{\mathrm{nd}}$ ed., Cambridge Univ. Press, Cambridge 2002.

\bibitem{allu1} D. V. Ahluwalia-Khalilova and D. Grumiller, Phys. Rev.
\textbf{D 72}, 067701 (2005) [{\tt arXiv:hep-th/0410192}].

\bibitem{n-j--l/1}
Y.~Nambu, G.~Jona--Lasinio,
\textit{Phys. Rev.} \textbf{122}, 345 (1961).









\bibitem{fab}
S.~Vignolo, L.~Fabbri and R.~Cianci,
J.\ Math.\ Phys.\ {\bf 52}, 112502 (2011) 
[{\tt arXiv:1106.0414 [gr-qc]}].










\bibitem{flags}
R. Penrose and W. Rindler, ``Two-Spinor Calculus and Relativistic Fields'',
Cambridge Univ. Press, Cambridge 1986; R. da Rocha, J. Vaz, Int. J. Geom. Meth. Mod. Phys. {\bf 4}, 547 (2007) [{\tt arXiv:math-ph/0412074}].

\bibitem{cra} J. P. Crawford, J. Math. Phys. \textbf{26},
1429 (1985) .

\bibitem{holl} P. R. Holland, {Found. Phys.} \textbf{16}, 708 (1986).



\bibitem{boehmergrafeno}
C.~G.~Boehmer and L.~Corpe, 
 J.\ Phys.\ A {\bf 45}, 205206 (2012) [{\tt arXiv:1204.0135 [math-ph]}].


\bibitem{Fabbri:2011kq}
L.~Fabbri, Gen. Rel. Grav. {\bf 45}, 1285 (2013) 
[{\tt arXiv:1108.3046 [gr-qc]}].


\bibitem{Fabbri:2012yg}
L.~Fabbri and S.~Vignolo, 
Int. J. Theor. Phys. {\bf 51}, 3186 (2012) [{\tt arXiv:1201.5498 [gr-qc]}].


\bibitem{FV1}
L.~Fabbri and S.~Vignolo, 
Class.\ Quant.\ Grav.\ {\bf 28}, 125002 (2011) 
[{\tt arXiv:1012.1270 [gr-qc]}].



\bibitem{Fabbri:2011ha}L.~Fabbri,
Phys.\ Lett.\ B {\bf 707}, 415 (2012) 
[{\tt arXiv:1101.1761 [gr-qc]}]; 
L.~Fabbri, 
[{\tt arXiv:1101.2334 [gr-qc]}]

\bibitem{Fabbri:2011mi}
L.~Fabbri,
Phys.\ Rev.\ D {\bf 85}, 047502 (2012) 
[{\tt arXiv:1101.2566 [gr-qc]}].

\bibitem{CCSV1}
S.~Capozziello, R.~Cianci, C.~Stornaiolo and S.~Vignolo,
Class.\ Quant.\ Grav.\ {\bf 24}, 6417 (2007) 
[{\tt arXiv:0708.3038 [gr-qc]}].

\bibitem{CCSV2}
S.~Capozziello, R.~Cianci, C.~Stornaiolo and S.~Vignolo, 
Int.\ J.\ Geom.\ Meth.\ Mod.\ Phys.\ {\bf 5}, 765 (2008) 
[{\tt arXiv:0801.0445 [gr-qc]}].

\bibitem{CV4}
S.~Capozziello and S.~Vignolo,
 Annalen Phys.\ {\bf 19}, 238 (2010)
 [{\tt arXiv:0910.5230 [gr-qc]}].

\bibitem{Rubilar}
G.~F.~Rubilar, 
 Class.\ Quant.\ Grav.\ {\bf 15}, 239 (1998).

\bibitem{Roc11}
R. da Rocha, A. E. Bernardini and J. M. Hoff da Silva,
\emph{JHEP} {\bf 04}, 110 (2011) [{\tt arXiv:1103.4759}].


\bibitem{Fabbri:2011mg}
L.~Fabbri,
Int. J. Theor. Phys. {\bf 52}, 634 (2013) [{\tt arXiv:1106.4695 [gr-qc]}].

\bibitem{Saha1}
B.~Saha and G.~N.~Shikin,
 J.\ Math.\ Phys.\ {\bf 38}, 5305 (1997) 
 [{\tt arXiv:gr-qc/9609055}].

\bibitem{Saha2}
B.~Saha and T.~Boyadjiev,
 Phys.\ Rev.\ D {\bf 69}, 124010 (2004) 
 [{\tt arXiv:gr-qc/0311045}].

\bibitem{t}
M.~Tsamparlis, {Phys. Lett. A} \textbf{75}, 27 (1979).

\bibitem{plbb} J.~M.~Hoff~da Silva and R.~da Rocha,
 Phys. Lett. B {\bf 718}, 1519 (2013) 
 [{\tt arXiv:1212.2406 [hep-th]}].

\bibitem{allu} D. V. Ahluwalia-Khalilova and D. Grumiller, 
 {}\emph{JCAP} \textbf{07}, 012 (2005) [\texttt{%
arXiv:hep-th/0412080}].





\bibitem{Basak:2012sn}
 A.~Basak, J.~R.~Bhatt, S.~Shankaranarayanan and K.~V.~P.~Varma, \emph{JCAP} {\bf 04} (2013) 025 
 [{\tt arXiv:1212.3445 [astro-ph.CO]}]; 
 H.~M.~Sadjadi,
 Gen.\ Rel.\ Grav.\ {\bf 44}, 2329 (2012) 
 [{\tt arXiv:1109.1961 [gr-qc]}].




\bibitem{sh}
I.~L.~Shapiro,
{Phys. Rept.} \textbf{357}, 113 (2002) [\texttt{arXiv:hep-th/0103093}].


\bibitem{op} V. Figueiredo, E. Capelas de Oliveira, W. A. Rodrigues Jr.,
 Int. J. Theor. Phys.
\textbf{29}, 371 (1990) .





\bibitem{where1} 
 R.~da Rocha and J.~M.~Hoff da Silva, Int.\ J.\ Geom.\ Meth.\ Mod.\ Phys.\ {\bf 6}, 461 (2009)
 [{\tt arXiv:0901.0883 [math-ph]}].
 
\bibitem{where2} R.~da Rocha and J.~G.~Pereira,
Int.\ J.\ Mod.\ Phys.\ D {\bf 16}, 1653 (2007) 
[{\tt arXiv:gr-qc/0703076 [gr-qc]}].
\end{thebibliography}
\end{document}